\documentclass[12pt,a4paper]{article}
\usepackage{epsfig}
\begin{document}

\begin{flushright}
{\bf FIAN/TD-09/04}\\{\bf ITEP-TH-35/04}
\end{flushright}

\begin{center}

{\bf On distribution of number of trades in different time windows in the stock market.}

\end{center}

\begin{center}

{\bf I.M.~Dremin$^{(a)}$, A.V.~Leonidov$^{(a,b)}$}

\end{center}

(a) {\it Theoretical Physics Department, P.N.~Lebedev Physics Institute, Moscow, Russia}

(b) {\it Institute of Theoretical and Experimental Physics, Moscow, Russia}

\bigskip

\begin{center}

{\bf Abstract }

\end{center}

Properties of distributions of the number of trades in different intraday time intervals for five stocks traded in
MICEX are studied. The dependence of the mean number of trades on the capital turnover is analyzed. Correlation
analysis using factorial and $H_q$ moments demonstrates the multifractal nature of these distributions as well as
some peculiar changes in the correlation pattern. Guided by the analogy with the analysis of particle multiplicity
distributions in multiparticle production at high energies, an evolution equation relating changes in capital
turnover and a number of trades is proposed. We argue that such equation can describe the observed features of the
distribution of the number of trades in the stock market.

\section{Introduction}

Statistical analysis of the stock market properties undertaken from the physicist's point of view has drawn a lot
of attention in the last decade. In fact, a large fraction of research in the new field of econophysics
\cite{MS,BP,HFF,V} is devoted to questions related to finance. One of the central issues in description of the
stock price dynamics is understanding the activity pattern characterized by the intensity of trading (number of
trades in some given time interval). In particular, the long-range volatility correlations (one of the most
important properties of the financial time series) are directly induced \cite{PGAGS00} by the long-range
correlations in trading activity discussed in \cite{BLM00}. Very recently the multiscaling of the stock market
activity was discussed, in the framework of a network approach to complex system dynamics, in \cite{EKYB04}.  The
analysis of the properties of the distribution of the number of trades has also allowed to prove that the
trade-generating process is of distinctly non-markovian nature, so that the time at which the new trade occurs
depends on the times of a lot of preceding trades \cite{L03,L04}. Therefore the study of higher-order correlations
is of great importance.

In the present paper we perform a detailed analysis of the distributions of the number of trades in a set of
chosen intraday time intervals. The technique used in our study is borrowed from the analysis of multiparticle
production in high energy physics, see, e.g., the review papers \cite{WDK96,DG01}. The analogy drawn is between
the distribution of the number of trades in some chosen temporal interval (finance) and the distribution of the
number of produced particles in some chosen rapidity (longitudinal momenta) interval, both referred to as
multiplicity distributions. This technique allows to study correlations in the systems of many trades (particles).
Both total and genuine (i.e., non-reducible to those of smaller groups of trades) correlations are considered for
sets of any number of trades. The analysis made in the context of multiparticle production has revealed a number
of remarkable properties of multiplicity distributions in different rapidity windows, in particular - their
multifractal nature and a special behavior of genuine correlations governed by a spectacular dynamics of the
particle production process. Therefore it is of obvious interest to repeat this analysis in the context of stock
price dynamics, which is a main issue addressed in the present paper. Its results inspired us to formulate a
dynamical model of trading activity in financial markets.

\section{Data processing}

The data we used in our analysis is the tick data for five stocks - EESR,MSNG, RTKM, SNGS and LKOH, traded at the
Moscow International Currency Exchange (MICEX) in the year of 2003. For our study we have chosen five time
intervals $\Delta T$: 5 min., 15 min., 45 min. and 495 min. (the trading day at MICEX).

Before turning to the analysis of the multiplicity distributions, let us discuss in some more details the
above-mentioned analogy between the trade-generating process in finance and multiparticle production in high
energy physics.

In high energy physics one studies the particle production at some fixed collision energy. The higher the energy,
the more particles are produced. In "financial" terms, one "invests" energy and gets particles from it. In
financial context we suggest therefore an analogy between the energy and the total capital turnover (inflow for
buy trades and outflow for the sell ones) in some chosen time interval. More precisely, we suggest an analogy
between the collision energy $E_{coll}=\sqrt{s}$\ (in the center of mass of colliding particles) and the capital
turnover $\cal J$. The total capital turnover in larger temporal intervals is thus disseminated into those in
smaller ones corresponding to particular distribution of capital in between the trades.

\begin{figure}[h]

\begin{center}
\epsfig{file=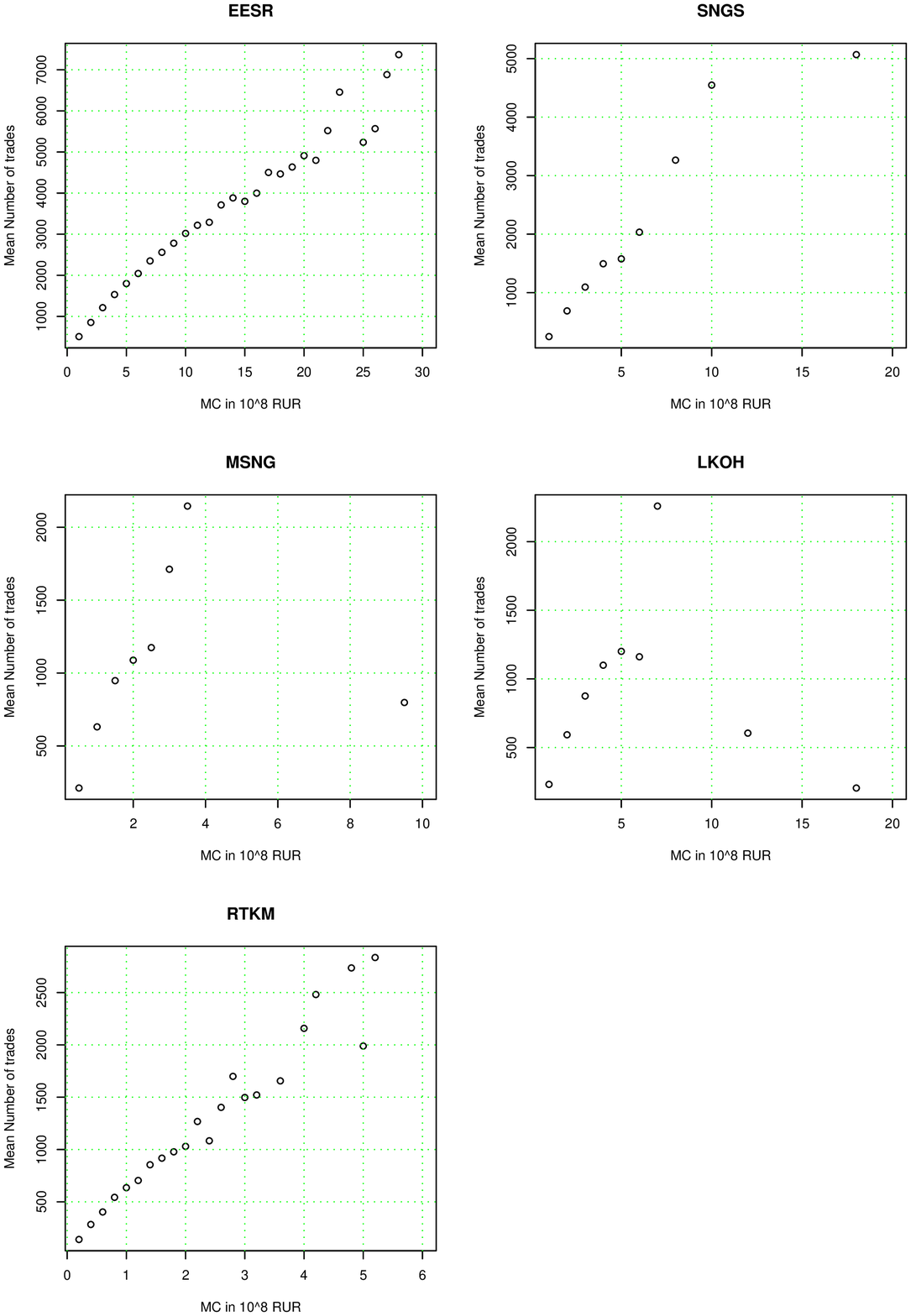,height=18cm,width=16cm}
\end{center}

\caption{Average multiplicity as a function of capital turnover, $\Delta T = 45$\ min.} \label{figmcntr}

\end{figure}

Let us illustrate the above-described analogy by considering the simplest characteristics relating the capital
turnover and trade multiplicity distributions  - a dependence of the mean number of trades in the time interval
under consideration $\langle n\rangle_{\Delta T}$ on the capital turnover $\cal J$$ _{\Delta T} $ (in
multiparticle dynamics the corresponding quantity is thus an energy dependence of mean multiplicity which has been
extensively studied, e.g., in \cite{DN94, DH93}). In Fig.~\ref{figmcntr} we show this dependence in the case of
$\Delta T = 45$\ min.

We see that the mean multiplicity is a concave function of the capital turnover ${\cal J}$, i.e. $\langle
N_{\Delta T} \rangle \propto {\cal J}^{\alpha}$ with $\alpha < 1$\footnote{In multiparticle dynamics this *would
correspond to $\langle n \rangle \sim s^{\alpha / 2}$ - a dependence characteristic to,* e.g., hydrodynamic models
of particle production}. For other time intervals the dependence is similar.

Equivalently, for the logarithmic turnover $y \propto \ln {\cal J}$,

\begin{equation}
 \langle n \rangle \propto {\rm exp} ( \alpha\cdot\,y ).   \label{multdep}
\end{equation}

\begin{figure}[h]

\begin{center}

\epsfig{file=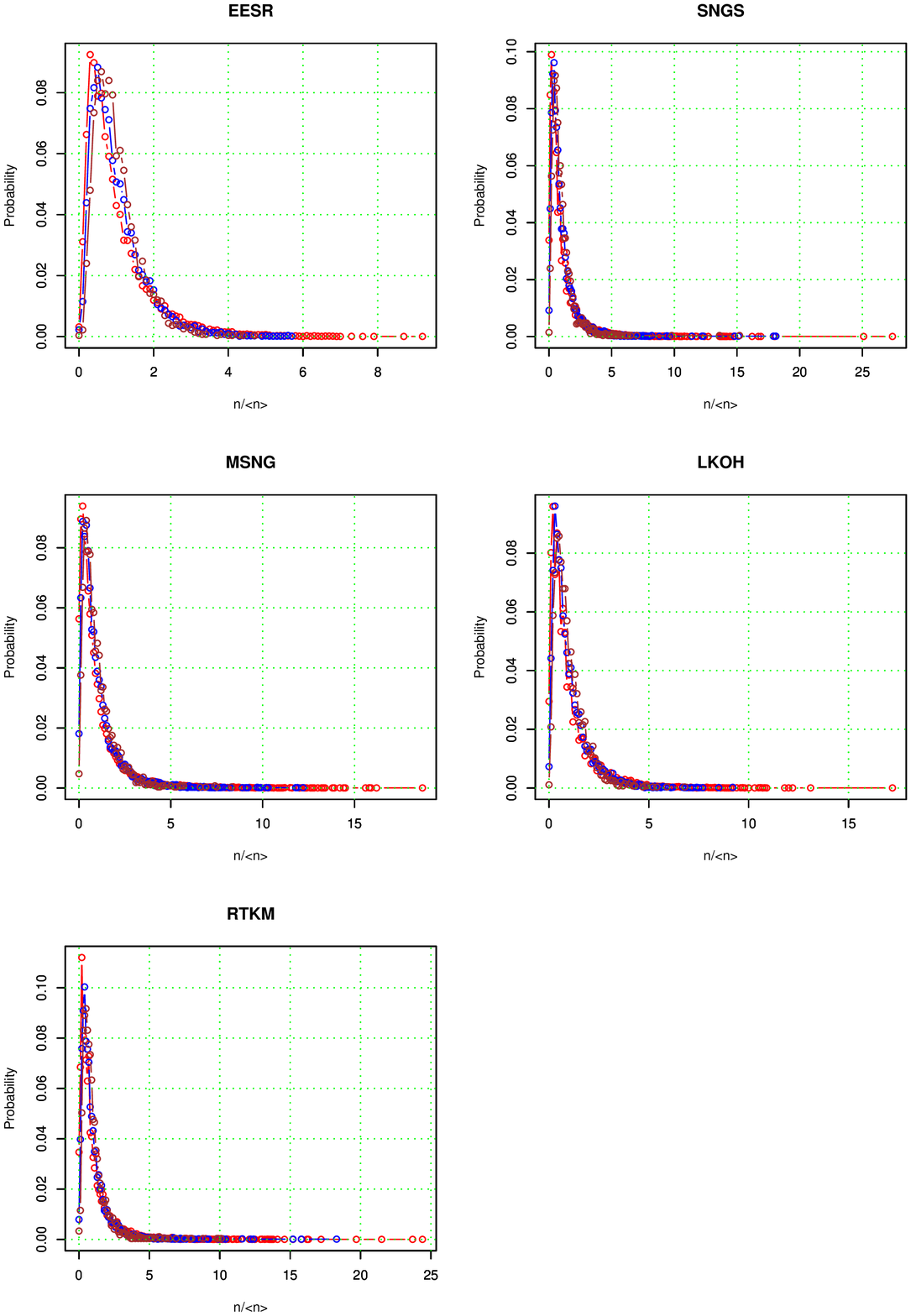,height=18cm,width=16cm}

\end{center}

\caption{Normalized multiplicity distributions, \, $\Delta T=$ 5 min (red), 15 min (blue); 45 min (brown); 495 min
(black)} \label{figdpdntot}

\end{figure}

Let us now turn to the analysis of other properties of the trade multiplicity distributions.  The corresponding
distributions (normalized at the mean number of trades in each interval) for the five stocks considered are shown
in Fig.~\ref{figdpdntot}.

 From Fig.~\ref{figdpdntot} we see, that normalizing the trade multiplicity distributions at the corresponding
averages brings them to the scale-invariant (i.e. independent of the size of the time window) form. Let us also
note that none of the standard distributions we have tried (Poisson, negative binomial, gamma, inverse gamma) give
a satisfactory fit of the data showing that we are dealing with some complex trade-generating dynamics.

The properties of the trade multiplicity distributions are conveniently summarized by their suitably defined
moments. Among the most important characteristics that are widely exploited, in particular, in analyzing the large
multiplicity events in multiparticle production at high energies, are factorial moments (normalized and
unnormalized), cumulants and their ratio, the so-called $H_q$ moments defined below.

The unnormalized factorial moments ${\cal F}_q$ with integer ranks $q \geq 1$ are defined by the following
formula:

\begin{equation}\label{facmom}
{\cal F}_q = \sum_n P(n) n (n-1) ... (n-q+1) = {d^q G(z) \over dz^q} \big |_{z=1},
\end{equation}
where $P(n)$ is the probability of having $n$ trades (the multiplicity distribution) for a given time interval.

The factorial moments (\ref{facmom}) are, by definiton, always positive. It is easily seen that the average
multiplicity is given by ${\cal F}_1$, the dispersion of the distribution is related to ${\cal F}_2$ etc. The
higher is the rank of the moment $q$, the more important is the contribution coming from the tail of the
distribution. They present therefore an ideal tool for focusing on the properties of the periods with intensive
trading.

In equation (\ref{facmom}) we have introduced a generating functional for the multiplicity distribution

\begin{equation}\label{genfun0}
G(z) \, = \, \sum_{n=0}^\infty z^n P(n).
\end{equation}

One often considers the normalized factorial moments

\begin{equation}\label{nfacmom}
F_q \equiv  {{\cal F}_q \over \langle n \rangle^q} \equiv {{\cal F}_q \over {\cal F}_1^q}.
\end{equation}

For Poisson distribution (independent trades) all normalized factorial moments are equal to 1. Thus a dependence
on the overall scale (mean multiplicity) is eliminated. Therefore the moments (\ref{nfacmom}) provide a convenient
basis for extracting information on the structure of correlations. Let us stress, that the factorial moments (both
normalized and unnormalized) are characterizing the overall correlation pattern, where the contributions of
genuine (irreducible) and reducible correlations are mixed. By genuine correlations we mean those in which all
subsets of the analyzed system are interconnected. The reducible ones appear in those systems which can be split
into two or more subsystems not connected to each other. The genuine irreducible correlations are described by the
cumulants ${\cal K}$:

\begin{equation}
{\cal K} _q \, = \,{d^q \ln G(z) \over dz^q} \big |_{z=1}. \label{cummom}
\end{equation}

For example, if only two-point correlations are present, all cumulants with ranks $q>2$ are zero, etc. In complete
absence of correlations (Poisson) all cumulants are zero.

\begin{figure}[h]

\begin{center}

\epsfig{file=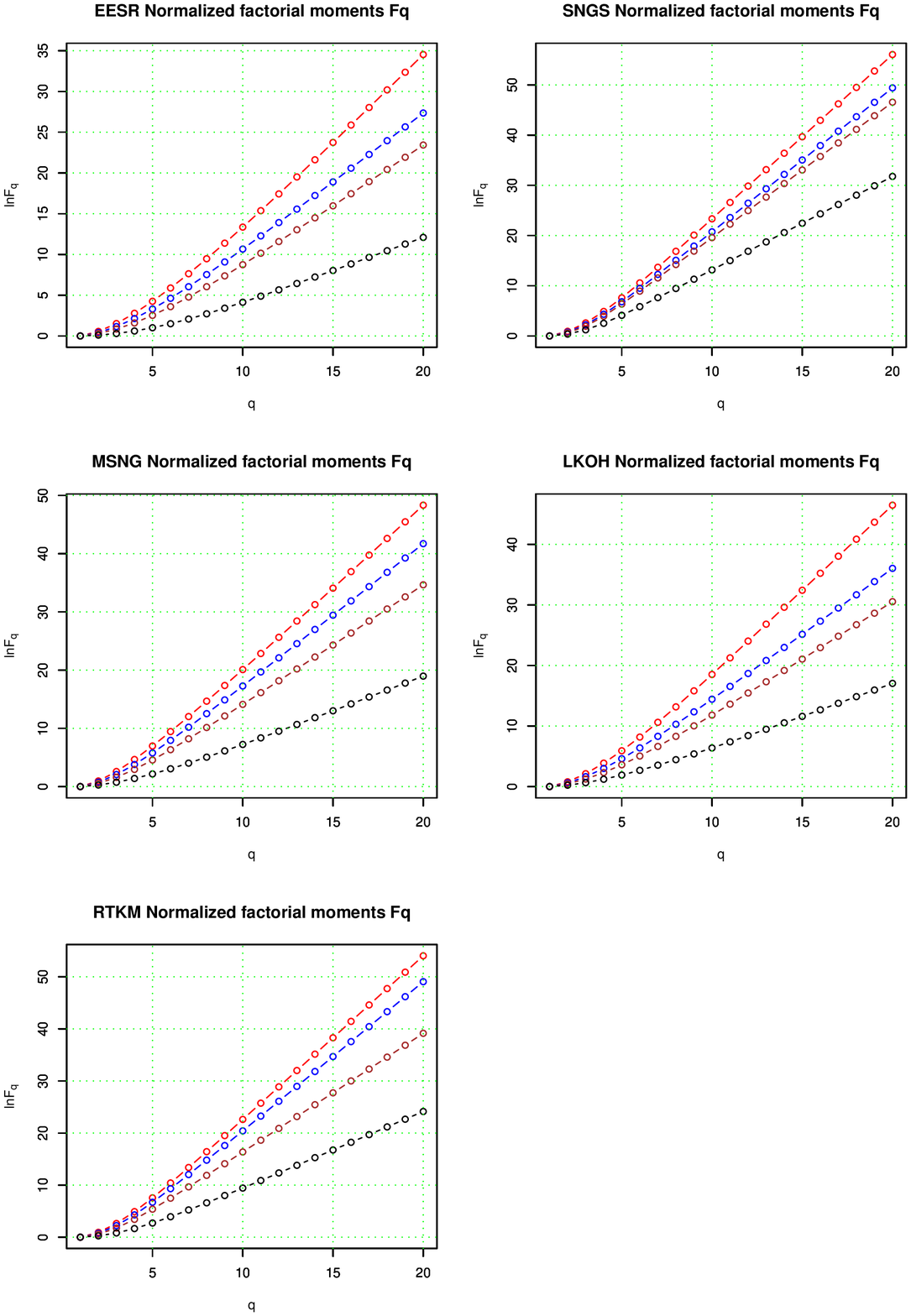,height=18cm,width=16cm}

\end{center}

\caption{Normalized factorial moments $F_q$;\, $\Delta T=$ 5 min (red), 15 min (blue); 45 min (brown); 495 min
(black).}

\label{fignfq}

\end{figure}

In what follows we shall focus ourselves on the properties of the normalized factorial moments $F_q$. In
Fig.~\ref{fignfq} we show the dependence of $F_q$ on the rank $q$.

We see, that, apart from the small bending at small $q$, the logarithms of the normalized factorial moments are
approximately proportional to the rank, $\ln F_q \propto q$, corresponding to an approximate exponential growth of
the factorial moments. Let us note in passing, that already this excludes the negative binomial distribution, for
which the growth of factorial moments with $q$ is faster than exponential.

\begin{figure}[h]

\begin{center}

\epsfig{file=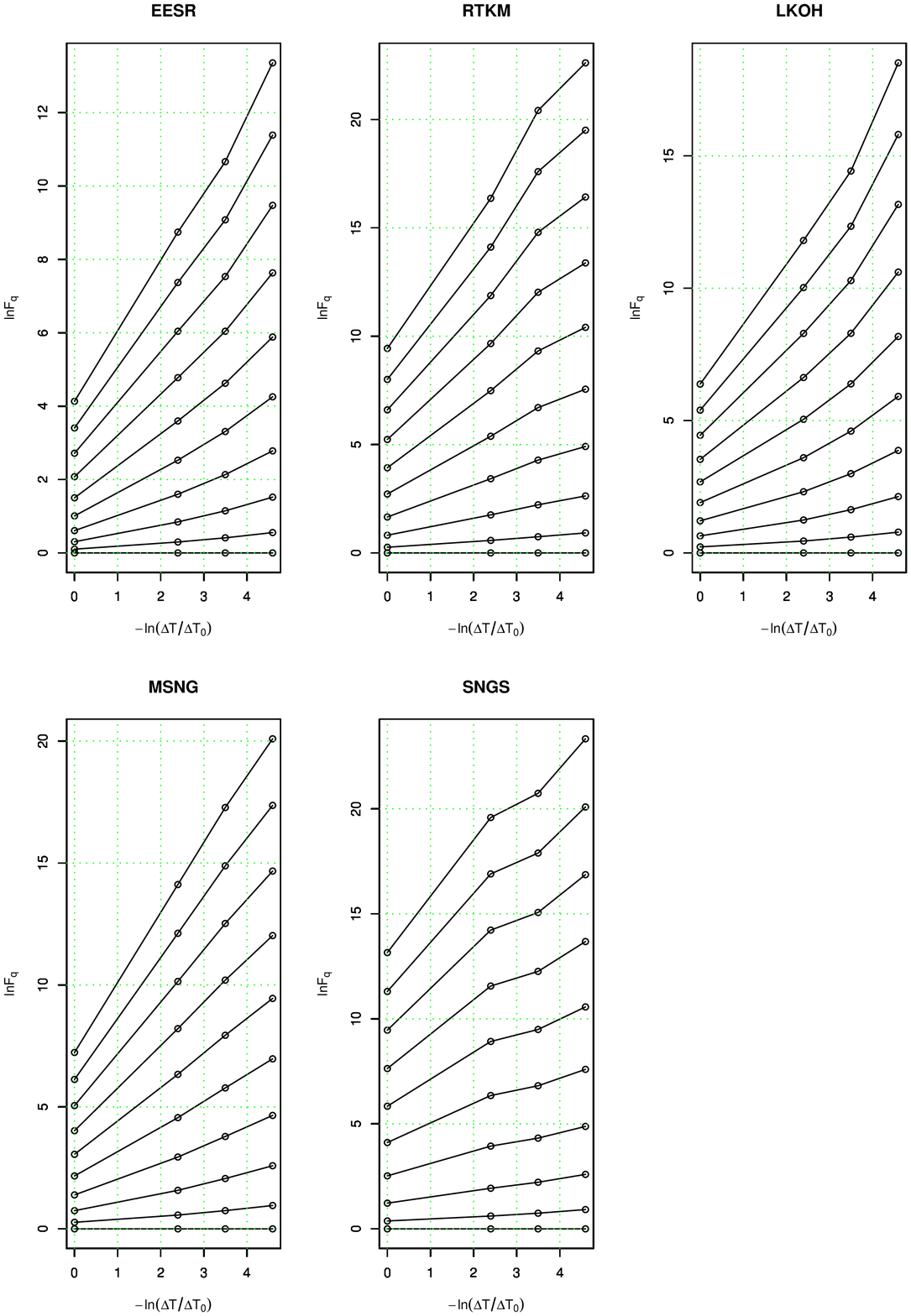,height=19cm,width=17cm}

\end{center}

\caption{{\small Normalized factorial moments $F_q$, $q=1$  (lowest line) $\cdots$ $q=10$ (highest line), as
functions of $\Delta T/\Delta T_0$; \, $\Delta T_0=495$\,min.}}

\label{fignfdelT}

\end{figure}

Of special interest is, of course, a dependence of the moments on the temporal resolution $\Delta T$, revealing
the scaling properties of the underlying trade multiplicity distributions. In particular, for generic fractal
system one has $\ln F_q (\Delta T) = \alpha_q \ln \Delta T$. If $\alpha_q \propto q$, the dynamics is of
monofractal nature. More complicated $q$-dependence of $\alpha_q$ indicates that we are dealing with a complex
multifractal process. The dependence of the normalized factorial moments $F_q$ on $\Delta T$ is shown, for the
moments with ranks in the interval $q=1 \cdots 10$, in Fig.~\ref{fignfdelT}.

\begin{figure}[h]

\begin{center}

\epsfig{file=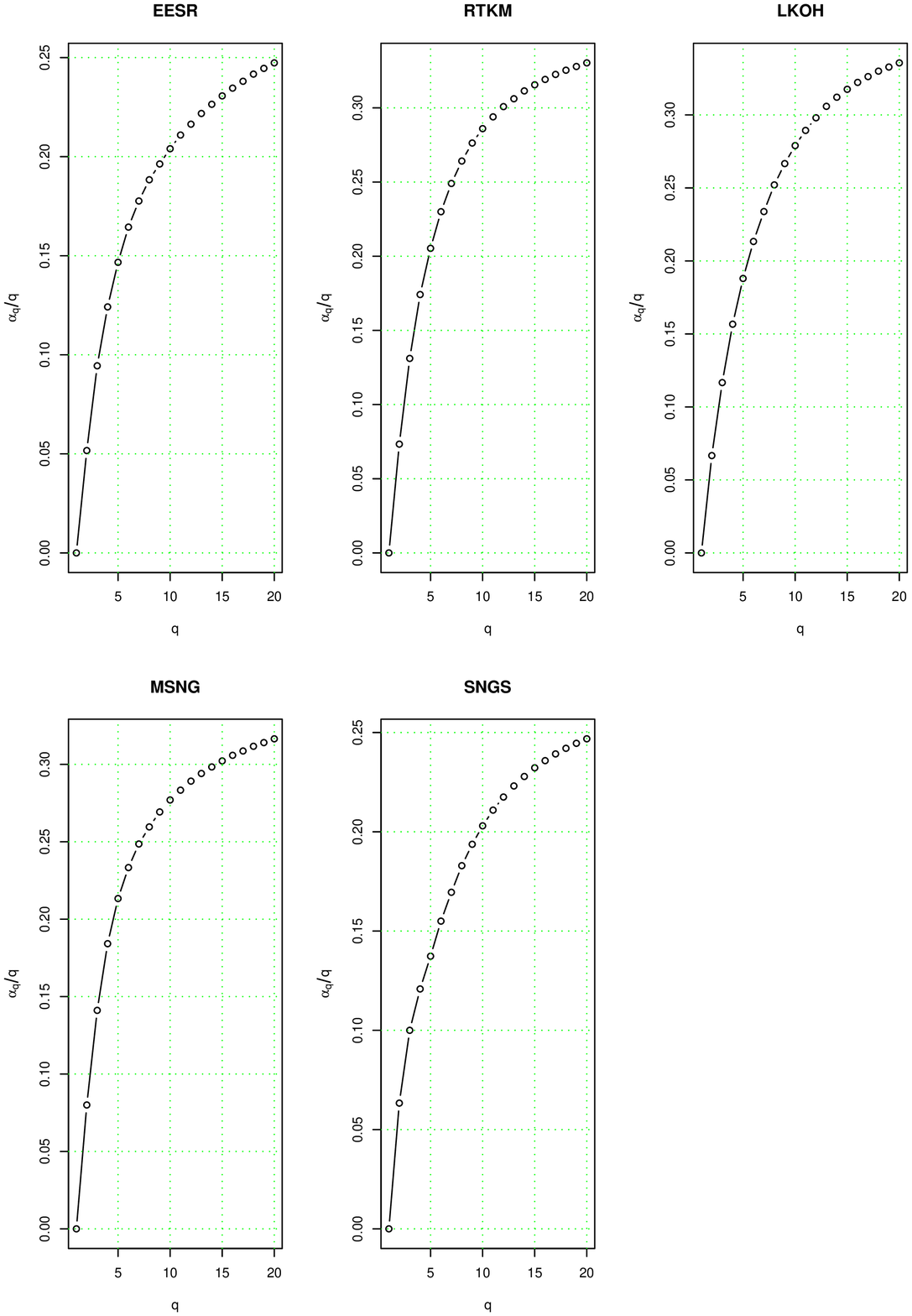,height=18cm,width=16cm}

\end{center}

\caption{Slopes of the normalized factorial moments $F_q$, $q=1 \cdots 20$.} \label{figslnfdelT}

\end{figure}

As mentioned above, the scaling pattern is described by the set of the slopes in Fig.~\ref{fignfdelT}. In
Fig~\ref{figslnfdelT} we show the ratio of the corresponding average slopes to the rank $q$, i.e., $\alpha_q / q$
.

The simple monofractal scaling would correspond to  $\alpha_q(q)/q \, =\,$ const., while from
Fig.~\ref{figslnfdelT} we clearly see that this is not the case thus pointing out to some complex multifractal
pattern. In Fig.~\ref{fignfdelT} we observe a characteristic strengthening of fluctuations (characterized by the
normalized factorial moments) with increasing resolution $\Delta T$. The similar feature is well known in quantum
chromodynamics \cite{DD93}\footnote{Multifractality of the price-generating process was discussed, in a different
context, in \cite{AMS98, MDB00, BDM01, BPM99, MAS03}.}.

As mentioned above, the factorial moments describe the overall (both reducible and irreducible) correlation
pattern. To characterize the irreducible content of the multiplicity distribution represented by cumulants, it is
convenient to consider the moments $H_q \equiv {\cal K}_q /{\cal F}_q$ which represent a share of genuine
correlations in their overall amount. They are more conveniently obtained from the factorial moments
(\ref{nfacmom}) by the recurrent relation

\begin{equation}\label{hqmom}
 H_q = 1 - \sum_{p=1}^{q-1} \frac {\Gamma (q)}{\Gamma (p+1) \Gamma(q-p)}
 \frac {F_p F_{q-p}}{F_q}H_{q-p}.
\end{equation}

\begin{figure}[h]

\begin{center}

\epsfig{file=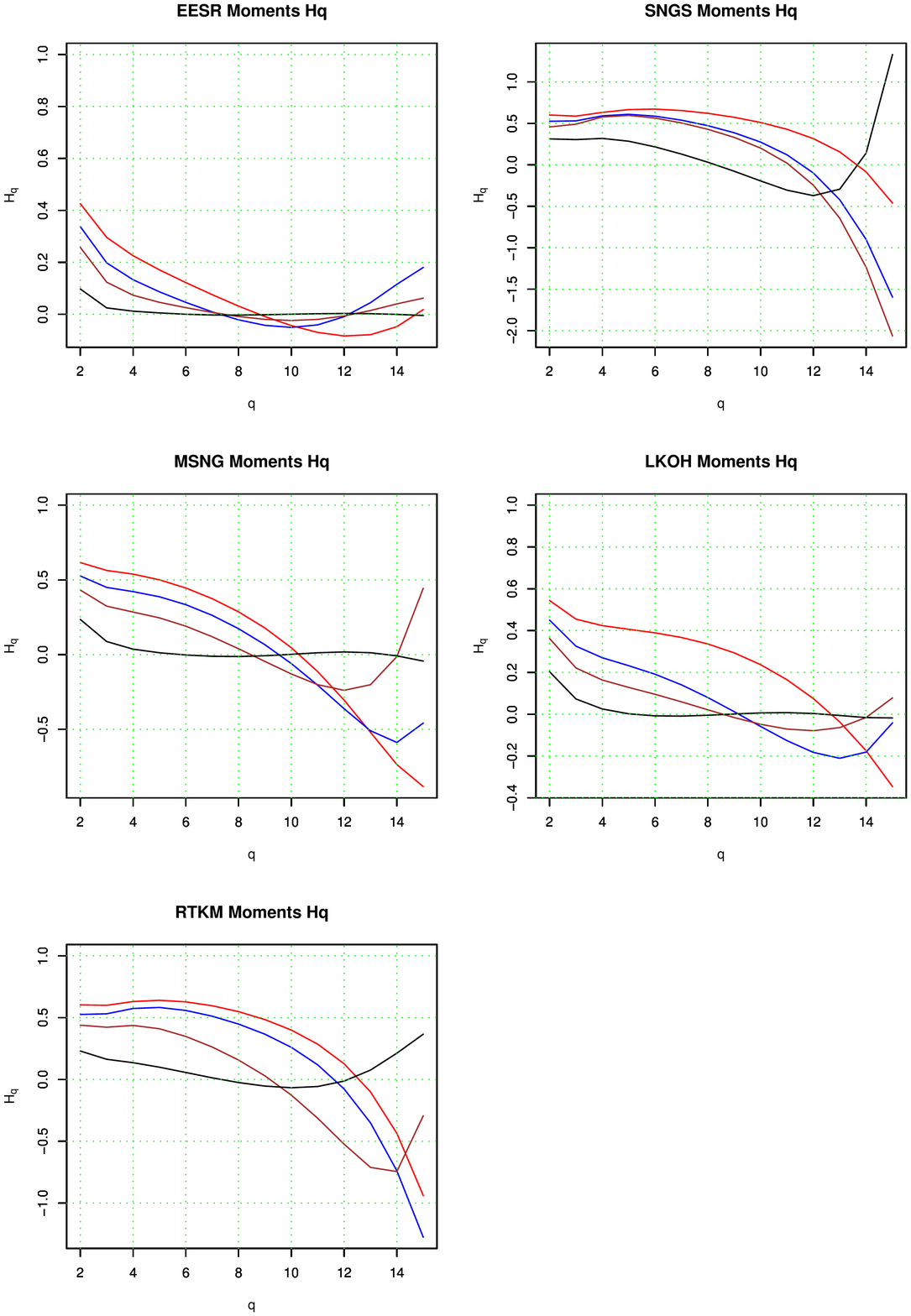,height=18cm,width=16cm}

\end{center}

\caption{Moments $H_q$\,,\,\,$\Delta T=$ 5 min (red), 15 min (blue); 45 min (brown); 495 min (black).}

\label{fighq}

\end{figure}
The moments $H_q$\,, plotted in Fig.~\ref{fighq}, show a very interesting behavior: they are changing sign at some
$q$ ! This phenomenon was first predicted theoretically in multiparticle dynamics \cite{D93} where it was used as
a tool allowing to uncover the details of the particle generation process. Afterwards it was confirmed by
experiment \cite{DA93, SLD}. Let us stress, that when working with "usual" distributions like Poisson, negative
binomial or gamma, the moments $H_q$ are non-negative. More exactly, they are zero for Poisson, positive for
negative binomial and non-negative for gamma.  To make it possible that $H_q$ become negative, one should
consider, at least, a mixture of these standard distributions, see \cite{IPPI}. Physically, positive cumulants
(and thus positive $H_q$) correspond to the effective repulsion between the points, and negative ones - to their
attraction (clustering of trades). Changing sign means passing through the "Poissonian" point, where the
corresponding irreducible contribution is absent. Below we shall consider a phenomenological dynamical model of
the capital turnover that allows to describe this phenomenon in the financial context.

\section{Mathematical model and its predictions}

The above analysis of the stock market data allows to formulate a mathematical model describing the trade
multiplicity distributions inspired by ideas originating from high energy physics. As already mentioned, the
method of correlation analysis discussed below was first suggested in \cite{D93} and is widely used in the studies
of multiparticle dynamics in high energy physics (see, e.g., a review article \cite{DG01}). In multiparticle
dynamics one considers multiplicity distributions in some region of phase space at given collision energy.
Particle creation is described within well-established and well-tested theory of strong interactions - quantum
chromodynamics. There exist sufficiently rigorous equations and well-defined predictions for the characteristics
of interest. Among them are those for the energy dependence of mean multiplicity, multifractal properties of the
particle production process and specific behavior of $H_q$ moments. We have observed the similar regularities in
the trade distributions analyzed above.

As has been already mentioned, in particular, the change in the sign of $H_q$-moments with growing
$q$ in multiparticle dynamics was theoretically predicted \cite{D93} and afterwards confirmed experimentally
\cite{DA93, SLD}. It has been extensively studied theoretically both for the electron-positron collisions
\cite{DN93, DH94, BFO} and for hadron-hadron ones \cite{IPPI}.
Let us note, that for standard probability distributions such as the negative
binomial one this effect does not take place. This points out to a quite
peculiar underlying dynamics both in multiparticle production and trading in
the stock market.

Let us try to write down phenomenological equations for the generating functional $G(z,y)$ for the probability
distribution of the trade multiplicity at given logarithmic turnover $y=\ln {\cal J}_{\Delta T}/{\cal J}_0$
(${\cal J}_0$= const):

\begin{equation}\label{genfun}
G(z,y) \, = \, \sum_{n=0}^\infty z^n P(n,y)
\end{equation}

In Eq.~(\ref{genfun}) the dependence on $y$ is shown explicitely (while it was not indicated in (\ref{genfun0})).

Following the analogy with quantum chromodynamics, we write

\begin{equation}\label{genfunev}
\frac {dG}{dy} = \gamma_0^2 \int_0^1 dx K(x) \bigl [ G(y+\ln x)G(y+\ln (1-x))-G(y) \bigr ].
\end{equation}

Equation (\ref{genfunev}) can be interpreted as follows. What we are trying to establish is a relation between the
distribution of trades in the intervals with differing capital turnover. The relation can be described as an
evolution from smaller to larger capital turnovers. It proceeds through adding  "extra" turnover, of fractional
size $x$, to the current one with fraction $1-x$ forming the total turnover given by $y$ - thus the quadratic
structure in the RHS of (\ref{genfunev}), describing the inflow to the total turnover. The linear term with
negative sign corresponds to the outflow from the total turnover and therefore it depends on $y$ only. The weight
of changing some fraction of the existing capital is specified by the kernel $K(x)$\footnote{Note that for
technical reasons it is convenient to work with the splitting variable $x$ denoting the relative yield of the
incoming capital in the newly formed (i.e., total) one.}. In equation (\ref{genfunev}) the constant $\gamma_0$
describes the strength of the overall evolution rate in a financial market and the integration is over the
splitting fraction $x$. The conservation of turnover is due to the arguments $\ln x$ and $\ln (1-x)$ in the right
hand side.

Let us assume that the kernel $K(x)$ has the following functional form:

\begin{equation}\label{kernel}
K(x) \, = \frac {1}{x} - c - d \, x,
\end{equation}
where $c$ and $d$ are constants. The $1/x$ term corresponds to the leading role of the trades having
parametrically small contribution to the capital turnover\footnote{In particle physics this corresponds to an
emission of particles having small energies - the so-called "infrared" particles.}. It is tempting to assume that
this effect can be related to the $1/f$ behavior found in the analysis of spectral densities of daily trades
\cite{BLM00}. The constant and linear terms describe the impact of large trades. In principle, next terms of the
Laurent expansion of the kernel can contribute as it happens in quantum chromodynamics. However, in
characteristics to be considered, it turns out that by keeping the $c$- and $d$ - terms in the kernel one can
conveniently parametrize all important effects. The equation (\ref{genfunev}) with the kernel (\ref{kernel}) can
be solved explicitely (see, e.g., \cite{DH93}). For the mean multiplicity the exponential dependence on $y$
follows directly if the coupling strength $\gamma _0$ does not depend on the capital turnover. Some non-linearity
in the exponent can be introduced by imposing such a dependence as it happens in quantum chromodynamics (see,
e.g., \cite{DN94}). The equation (\ref{genfunev}) also reveals the multifractal properties of the distributions
\cite{DD93} noticed by us in the above analysis.

To proceed with $H_q$-moments, let us rewrite equation (\ref{genfun}) as an expansion in the normalized factorial
moments $F_q$:

\begin{equation}\label{genfun1}
 G(z,y) \, = \, \sum_{q=0}^{\infty} z^q \langle n \rangle^q F_q.
\end{equation}

Assuming that the dependence on the capital turnover $y$ is coming only through the average multiplicity $\langle
n \rangle$

\begin{equation}
 \langle n \rangle \propto \exp (\gamma y),
\end{equation}

we obtain the following equation for the factorial moments\footnote{Technically the condition of the approximate
$F$-scaling $F'_q \ll \gamma q F_q$ should hold.}:

\begin{equation}\label{fmev}
 \gamma q F_q = \gamma_0^2 \int_0^1 dx K(x) \left [ \sum_{l=0}^q x^{\gamma l} (1-x)^{ \gamma(q-l)}
 F_l F_{q-l} - F_q \right ].
\end{equation}

Equation (\ref{fmev}) is solved iteratively starting from $q=1$. At $q=1$ one gets the following equation for the
rate $\gamma$ of the growth of mean multiplicity with increasing capital turnover ($F_0=F_1=1$):

\begin{equation}\label{fmev1}
 \gamma = \gamma_0^2 \int_0^1 dx K(x) \left [x^\gamma + (1-x)^\gamma -1 \right ].
\end{equation}

 From (\ref{fmev1}) one can find  $\gamma$ as a function of $\gamma_0$ and $c$.
With good accuracy the corresponding equation can be reduced to an algebraic
one:

\begin{equation}\label{fmev2}
  \gamma^2 = \gamma_0^2 \left [
 1 + \left ( C-c-\frac{d}{2} \right )\gamma -
 \left (\frac{\pi^2}{6} -2c -\frac{d}{2} \right ) \, \gamma^2 \right ]\,,
\end{equation}
where $C\approx 0.577$ is Euler constant. At $\gamma \ll 1$ one has $\gamma \approx \gamma_0$. More accurately,
$$
\gamma \approx \gamma_0 \left [1+ \frac{1}{2} \left ( C -c -\frac{1}{2} d \right ) \gamma_0 \right ]
$$
The magnitude of $\gamma$ determines the growth of the average trade multiplicity $\langle n \rangle \propto {\rm
exp}(\gamma y)$ with $y$. In the weak coupling regime $\gamma$ is practically equal to $\gamma_0$.

To understand the behavior of the moments $H_q$, let us first consider the case of $c=d=0$. Then the contribution
to the RHS of the evolution equation (\ref{fmev}) is dominated by small $x$, and $G(y+\ln (1-x))$ can simply be
replaced by $G(y)$, so that

\begin{equation}\label{fmev3}
 \left ( \ln G \right )' = \gamma_0^2 \int_0^1 \frac {dx}{ x} \left [ G(y+\ln x)-1 \right ] =
 \gamma_0^2 \int_{-\infty}^y dz [G(z)-1].
\end{equation}

Differentiating both sides of (\ref{fmev3}) over $y$, we have

\begin{equation}\label{fmev4}
 \left ( \ln G \right )'' \, = \, \gamma_0^2 [G-1].
\end{equation}

Using Eqs (\ref{facmom}), (\ref{cummom}), one gets at $\gamma \approx \gamma_0$
\begin{equation}\label{hqm}
 H_q \, = \, \frac {\gamma_0^2}{\gamma^2 q^2} \approx \frac{1}{q^2}.
\end{equation}

From Eq.~(\ref{hqm}) we conclude that the $H_q$-moments are always positive if there are small contributions to
the capital turnover only.

Taking into account the remaining terms in the kernel $K(x)$, we get

\begin{equation}\label{hfmev5}
 H_q \, = \, \frac {\gamma_0^2}{\gamma^2 q^2} \left ( 1+\left(C-c-\frac{1}{2} d \right) q \right ).
\end{equation}

Now, $H_q$ change sign at some value of $q$. We see, that the zero of the $H_q$ moments is located at
 $q=\left(c+\frac{1}{2} d -C\right)^{-1}$.
At larger $q$ the moments $H_q$ are negative and reach their minimum at some $q_{min}$. The larger is the
contribution of "strong" dealers (large $c$ and (or) $d$) the lower is the intercept of $H_q$ with the $q$-axis.
In our analysis of stock data we have just met with such a behavior. Let us also note, that from
Fig.~(\ref{fighq}) we see, that the data with lower two-point ($H_2$) correlation have a smaller intercept of
$H_q$. This implies that "strong" dealers are more influential in markets (or time intervals) characterized by
weak interaction with other market participants. Various stocks differ in their relative roles - as seen from
comparison of various plots in Fig.~(\ref{fighq}). It would certainly be of interest to compare these intercepts
for different markets. This is an important signature of long-range correlations involved.

To make the above analysis more exact, a numerical recursive solution of (\ref{fmev}) is required. For example, at
$q=2$ we get the solution in the form

\begin{equation}\label{fmev5}
 F_2 = \frac {\int_0^1 dx K(x) x^\gamma (1-x)^\gamma }{ 2 \int_0^1 dx K(x) \left [x^\gamma+(1-x)^\gamma -1 \right ] -
 \int_0^1 dx K(x) \left [x^{2\gamma}+(1-x)^{2\gamma} -1 \right ]} .
\end{equation}

In the above qualitative argumentation we have used, following the analogy with QCD, the small - $\gamma$
approximation. To get a realistic description of financial data in question one has to perform a deeper analysis
of the formulae describing, e.g., the $H_q$ moments in the suggested formalism. At present, we limit ourselves by
these qualitative statements. Quantitative estimate of theoretical parameters from the data and its implications
will follow.

\section{Conclusions}

We have analyzed the tick data for five stocks traded at MICEX and found some
common qualitative features for all of them. The most important ones are:
\begin{enumerate}

\item{The normalized trade multiplicity distributions are, to a good accuracy, scale-invariant with respect to the
size of the time window considered.}

\item{The average multiplicity of trades is a concave function of the total capital
turnover.}

\item{The multiplicity distributions demonstrate a multifractal behavior.}

\item{The cumulant (and therefore $H_q$) moments change their sign at some rank.}
\end{enumerate}

These features are strongly reminiscent of the properties of particle multiplicity distributions studied in high
energy physics where the well-developed theory of quantum chromodynamics describes (and moreover predicts) them
quite successfully. Using this analogy, we have proposed the phenomenological model of trades multiplicity
distribution with the equation similar to that arising in quantum chromodynamics. This equation provides a
qualitative description of the above-mentioned properties. Its interpretation in terms of inflow and outflow of
the capital turnover, providing some insight into the dynamics of financial markets, is considered.

The work was supported by RFBR grants 04-02-16880, 04-02-16445-a, 02-02-16779 and the Scientific Schools Support
Grant 1976.2003.02.

\end{document}